\documentclass[aps,prl,showpacs,twocolumn,groupedaddress]{revtex4}

\usepackage{graphicx}
\usepackage{amssymb}
\usepackage{color}

\newcommand{\rs}{\rm \scriptscriptstyle}

\begin{document}

\title{Repulsive shield between polar molecules}

\author{A.~V.\ Gorshkov$^1$}
\author{P. Rabl$^{1,2}$}
\author{G.\ Pupillo$^{3,4}$}
\author{A.\ Micheli$^{3,4}$}
\author{P.\ Zoller$^{3,4}$}
\author{M.~D.\ Lukin$^{1,2}$}
\author{H.~P.\ B\"uchler$^5$}

\affiliation{$^1$Physics Department, Harvard University, Cambridge, Massachusetts 02138, USA}
\affiliation{$^2$Institute for Theoretical Atomic, Molecular and Optical
Physics,
Cambridge, MA 02138, USA}
\affiliation{$^3$Institute for Quantum Optics and Quantum Information, 6020
Innsbruck, Austria}
\affiliation{$^4$Institute for Theoretical Physics, University of Innsbruck, 6020 Innsbruck,
Austria}
\affiliation{$^5$Theoretische Physik III, University of Stuttgart, 70155
Stuttgart, Germany}

\date{\today}

\begin{abstract}

We propose and analyze a technique that allows to suppress inelastic
collisions and simultaneously enhance elastic interactions between cold
polar molecules. The main idea is to cancel the leading dipole-dipole
interaction with a suitable combination of static electric and microwave
fields in such a way that the remaining van-der-Waals-type
potential forms a three-dimensional repulsive shield. We analyze
the elastic and inelastic scattering cross sections relevant 
for evaporative cooling of polar molecules and discuss the prospect for
the creation of crystalline structures.
%
\end{abstract}

\pacs{34.50.Rk, 37.10.Mn, 64.70.dg}





\maketitle

It is well known that an attractive van der Waals interaction dominates the
microscopic interaction potential between ground state atoms and molecules at
short distances \cite{weiner99}. As a consequence,  ultracold atomic and molecular gases are
prone to the formation of deeply bound molecular states via inelastic
collisions.  
These inelastic collisions limit the lifetime of strongly
interacting ultracold gases \cite{weiner99, bohn02}. Controlling inelastic collisions is therefore
crucial for the creation of novel strongly correlated quantum degenerate
systems, such as polar molecules.  In this Letter, we demonstrate the
possibility to engineer a three-dimensional repulsive interaction between polar
molecules, which allows for the suppression of inelastic collisions, while
simultaneously enhancing elastic collisions. This 
technique may open
up a way towards
efficient evaporative cooling and the creation of novel long-lived quantum
degenerate gases of polar molecules.

A special property of polar molecules prepared into the lowest rotational and
vibrational state is a permanent electric dipole moment $d$, which gives rise to
tunable dipole-dipole interactions and offers a large potential for the creation
of strongly correlated quantum phases
\cite{goral02,micheli06,rezayi05,buechler07.2}.  Two routes are currently
explored for the experimental realization of quantum degenerate polar molecules:
(i) trapping and cooling of preexisting molecules
\cite{weinstein98,tarbutt04,maxwell05,meerakker05,campbell07,sawyer07} and  (ii)
synthesizing polar molecules from a cold mixture of atomic gases
\cite{bahns00,dion01,jones06,kerman04,wang04,haimberger04,kraft06}.  While
scattering properties of heteronuclear molecules with dipole-dipole interactions
are currently theoretically explored \cite{avdeenkov06,ronen06,krems05}, it is
expected that inelastic collisions strongly increase for polar molecules
compared to atomic systems due to the opening of additional decay channels.

\begin{figure}[htb]
\begin{center}
\includegraphics[width= 0.9\columnwidth]{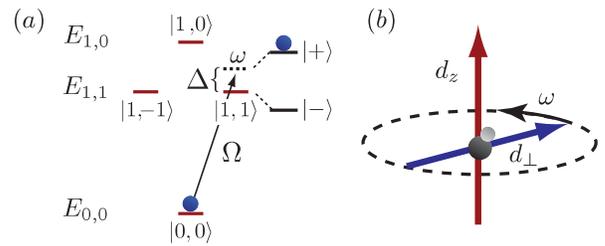}
\end{center}
\caption{(a) Energy levels of $H_{\rs rot}^{(i)}$: 
the arrow indicates  the microwave field.  
(b) Sketch for the cancelation
of the dipole-dipole interaction: $d_{z}$ denotes the dipole moment induced by 
 the static electric field, while the dipole moment $d_{\perp}$ is
rotating in an orthogonal plane due to the microwave coupling. } \label{Fig1}
\end{figure}

The main idea of our approach 
is to cancel the leading dipole-dipole interaction with a
suitable combination of a static electric field and a continuous-wave microwave
field: the former induces a dipole moment $d_{z}$, while the latter drives an
additional dipole moment $d_{\perp}$ rotating with frequency $\omega$ of the
microwave field, see Fig.~\ref{Fig1}(b). In analogy with magic-angle techniques
in NMR \cite{slichter96}, the time-averaged interaction of the rotating dipole
moment shows  a negative sign allowing for a cancelation of the total
dipole-dipole interaction. The remaining
interaction is tunable in strength with a repulsive van der Waals behavior
$V_{\rs eff} \sim (d^4/\hbar \Delta)/r^6$, where $\Delta$ is the detuning of the
microwave field and $r$ is the intermolecular separation.
The three-dimensional shield described here is thus purely repulsive. 
This is in contrast to the ``blue  shield'' discussed in the
context of atomic gases, which is attractive at large distances \cite{weiner99}.
We find that the efficiency of the shield is
determined by a single dimensionless parameter $\gamma = d^2 m/\hbar^2 r_{\rs
B}$ with $r_{\rs B} = (d^2/B)^{1/3}$, $B$ the rotational energy, and $m$ the
mass of the molecule. 
Under optimal conditions, for large values of $\gamma$, inealstic collisions can be 
quenched for temperatures $T \lesssim 0.01 B$.


%
%

To describe the shield quantitatively, we consider
two polar molecules prepared into the
vibrational ground state. The interaction at large intermolecular distances 
is dominated by the dipole-dipole interaction, yielding the Hamiltonian
%
%
%
%
\begin{equation}
  H = \frac{{\bf P}^2}{4m} + \frac{{\bf p}^2}{m} 
  + \frac{{\bf d}_{1} {\bf d}_{2} 
  - 3 ({\bf d}_{1}  \hat{{\bf r}}) ({\bf d}_{2}  \hat{{\bf r}})}{r^3} 
  + \sum_{i=1}^2 H_{\rs rot}^{(i)},
 \label{PMHamiltonian}
\end{equation}
where we have introduced the center of mass ${\bf R}=\left({\bf r}_{1}+{\bf
r}_{2}\right)/2$ and the relative coordinate ${\bf r}= {\bf r}_{1}-{\bf r}_{2}$,
with the corresponding momenta ${\bf P}$ and ${\bf p}$, respectively, 
while $r = |{\bf r}|$ and $\hat{{\bf r}}={\bf r}/r$. 
Here, we are interested in polar molecules in the $^1\Sigma$  electronic
ground state, and the Hamiltonian for the internal structure  takes the form
\cite{micheli07}
\begin{equation}
  H_{\rs rot}^{(i)} = B \: {\bf J}_{i}^2 - {\bf d}_{i} \cdot {\bf E}_{\rs dc} -
  {\bf d}_{i} \cdot {\bf E}_{\rs ac}(t) 
\end{equation}
with the dipole operator ${\bf d}_{i}$ and the permanent dipole moment $d$.  The
first term describes a rigid rotor accounting for rotational structure with the
rotational energy $B$, while the last two terms describe the coupling to an
external static electric field ${\bf E}_{\rs dc}$ and microwave field ${\bf
E}_{\rs ac}$. While additional interactions with the nuclear spins are in
general small and can be ignored, the analysis presented here remains valid for
polar molecules with an electronic spin  in a strong magnetic field with the
Zeeman splitting larger than the energy scales of the shield and the
spin-rotation coupling; such a situation naturally appears for some polar
molecules in magnetic traps.

We choose to apply a static electric field ${\bf E}_{\rs dc} = E_{\rs dc} {\bf
e}_{z}$  along the $z$-axis.  For each molecule, a
suitable basis set for the internal states is given by the eigenstates of the
rotor Hamiltonian in the external static field.  These states and the
corresponding energies depend on the dimensionless parameter $d E_{\rs dc}/B$
and are denoted by  $|j, m_z \rangle_{i}$ and $E_{j,m_z}$, respectively, with
$m_z$ the angular momentum along the $z$-axis and $j$ denoting the different
energy manifolds, see Fig.~\ref{Fig1}(a). In addition, we apply a 
circularly
($\sigma_+$) polarized microwave field ${\bf E}_{\rs ac}(t)$ 
propagating along the $z$-axis and coupling dominantly the ground state
$|0,0\rangle_{i}$ with the first excited state $|1,1\rangle_{i}$. The microwave
field is characterized by the detuning $\Delta = \omega-
(E_{1,1}-E_{0,0})/\hbar$ and Rabi frequency  $\Omega =  E_{\rs ac} d_{c}/\hbar$
with the dipole coupling $ d_{c} = |\langle 0,0|{\bf d}_{i}|1,1\rangle|$.  The
leading effect of the microwave field on the internal structure of each molecule
is to 
mix 
the ground state $|0,0\rangle$ with the excited state $|1,1\rangle$.
We are interested in the regime with $\Delta,\Omega \ll B/\hbar$ and $d E_{\rs
dc}< 2 B$, where the rotating wave approximation is valid. In the rotating
frame, these dressed states then take the form $|+\rangle= \alpha |0,0\rangle +
\beta |1,1\rangle$ and $|-\rangle= \beta |0,0\rangle - \alpha |1,1\rangle$ with
the energy splitting $\Delta E = \hbar \sqrt{\Delta^2+ 4 \Omega^2}$ and $\alpha
= - A/\sqrt{A^2+  \Omega^2}$, $\beta=  \Omega/\sqrt{A^2+ \Omega^2}$, and $A =
(\Delta +\sqrt{\Delta^2 + 4 \Omega^2})/2$.  Throughout this letter, we are
interested in a shield with a high barrier, which is optimized for 
parameters close to $d E_{\rs dc}/B = 1$, $\hbar \Delta = 0.015 B $, and
$\Omega/\Delta = 0.9258 $ \footnote{For ${\rm Li Cs}$ the DC fields $\sim\! \!2
{\rm kV/cm}$ and microwave intensities  $\sim \! 10 {\rm W/cm^2}$ are achievable
in the laboratory.} (see below).


%
\begin{figure}[htb]
\begin{center}
\includegraphics[width= \columnwidth]{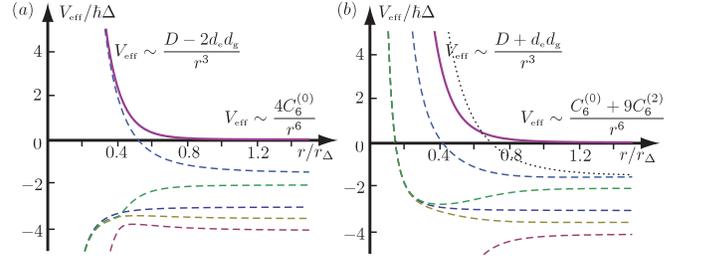}
\end{center}
\caption{Born-Oppenheimer potentials in the limit $r \gg r_{\rs B}$: (a)
$\theta =0$ and (b) $\theta= \pi/2$.  The effective potential  $V_{\rs eff}({\bf
r})$ (solid line) is repuslive for all angles $\theta$. The 
dotted line denotes the antisymmetric level
relevant during a three-body collision.}  \label{Fig2}
\end{figure}

We next turn to the dipolar interaction and derive the dressed Born-Openheimer
potentials. Each polar molecule is prepared into the internal state
$|+\rangle_{i}$ by an adiabatic switching on of the microwave field.
Consequently, the effective interaction potential $V_{\rs eff}({\bf r})$ is
determined by the dressed Born-Oppenheimer potential adiabatically connected to
the state $|+\rangle_{1}|+\rangle_{2}$, see Fig.~\ref{Fig2}.
%
%
The competition between the rotational splitting $B$ and the dipole-dipole
interaction provides a characteristic length scale $r_{\rs B} = (d^2/B)^{1/3}$.
For large interparticle distances $r \gg r_{\rs B}$, the dipole-dipole
interaction is weak and does not couple different rotor levels.  Consequently,
the only relevant coupling appears due to the microwave field bewteen the
manifolds with $j=0$ and $j=1$. The relevant internal states are then given by
the three states  $|g\rangle_{i}=|0,0\rangle_{i}$,
$|e\rangle_{i}=|1,1\rangle_{i}$, and $|\bar{e}\rangle_{i}=|1,-1\rangle_{i}$.
As the microwave field couples each polar molecule with the same phase, the
Born-Oppenheimer potentials separate into $6$ symmetric and $3$ antisymmetric
dressed potentials. The effective potential adiabatically connected to the state
$|+\rangle_{1}|+\rangle_{2}$ is symmetric, and therefore, we can restrict the
analysis to the symmeric potentials: a basis is given by the symmetric states
$|g,g\rangle$, $|e, g\rangle$, $|e,e\rangle$, $|g,\bar{e}\rangle$,
$|e,\bar{e}\rangle$, and $|\bar{e},\bar{e}\rangle$.  Within the rotating frame,
the Hamiltonian projected onto this subspace reduces to
\begin{equation}
  H=\left( 
  \begin{array}{cccccc}
  d^{2}_{g} \nu & \sqrt{2}\hbar \Omega  &0 &0 & 0& 0\\
  \sqrt{2}\hbar \Omega & H_{eg} &\sqrt{2}\hbar \Omega & d_{c}^2 \mu^*/2 &0& 0\\
0 & \sqrt{2}\hbar \Omega & H_{ee}  & 0 & 0 & 0\\
0 & d_{c}^2 \mu/2 & 0 & H_{eg} &\hbar \Omega & 0\\
0& 0& 0& \hbar \Omega & H_{ee} & 0\\
0 & 0 & 0 & 0 & 0 &H_{ee}
\end{array}\right)
\label{symmetricdipole}
\end{equation}
with the dipole moments  $d_{g}=|\langle g|{\bf d}_{i}| g\rangle|$,
$d_{e}=|\langle e|{\bf d}_{i}| e\rangle|$,  and $H_{eg}= (d_{e} d_{g}-
d_{c}^2/2) \nu -\hbar \Delta$ and $H_{ee} = d_{e}^2 \nu - 2 \hbar \Delta$.  The
terms $\nu = (1-3\cos^2\theta)/r^3$ and $\mu = 3 \sin^2 \theta e^{2 i \phi}
/r^3$ describe the spatial dependence of the dipole-dipole interaction, with
$\theta$ and $\phi$ being the polar and azimuthal angles of ${\bf r}$,
respectively.  The Born-Oppenheimer potentials then follow from a
diagonalization of the Hamiltonian $H$ and are shown in Fig.~\ref{Fig2}, with
the level adiabatically connected to the state $|+\rangle_{1} |+\rangle_{2}$
(solid line) giving rise to the effective interaction $V_{\rs
eff}({\bf r})$.

The detuning $\Delta$ introduces a new length scale in the problem $r_{\Delta} =
(d^2/\hbar \Delta)^{1/3} \gg r_{\rs B}$, e.g., for LiCs with $B/\hbar\approx 5.8
{\rm GHz}$, $r_{\rs B}\approx 9.2$ nm and $r_\Delta \approx 37.5 {\rm nm}$ at
$\hbar \Delta = 0.015 B$. At large interparticle distances $r > r_{\Delta}$, the
Born-Oppenheimer potentials are well described by perturbation theory in the
dipole-dipole interaction. The static electric field gives rise to a dipole
moment $d_{z}  = \alpha^2 d_{g} + \beta^2 d_{e}$ along the $z$-axis, while the
microwave field drives an additional dipole moment $d_{\perp} = \sqrt{2} \alpha
\beta d_{c}$ rotating with frequency $\omega$ in the $x$-$y$ plane.  The
combination of the two dipole forces provides the interaction $V_{\rs eff}({\bf
r})= (d_{z}^2-d_{\perp}^2/2) \left(1-3 \cos^2\theta\right)/r^3$. A proper choice
of the two parameters $E_{\rs dc} d /B$ and $\Omega/\Delta$ gives
$d_{z}=d_{\perp}/\sqrt{2}$, providing a cancelation of the leading dipole-dipole
interaction \cite{buechler07.2}.  The remaining interaction then follows from
second order perturbation theory and provides a van-der-Waals-type repulsion
\begin{equation} 
V_{\rs eff}({\bf r}) = \frac{1}{r^6} \left[C_{6}^{(0)} (1-3
\cos^2 \theta)^2 + C^{(2)}_{6}\: 9 \sin^4 \theta\right]
\label{largeRpot}
\end{equation}
with
\begin{eqnarray}
\hbar C_{6}^{(0)} & = & \frac{\alpha^2\beta^2}{ \sqrt{\Delta^2 + 4 \Omega^2}} \left\{\frac{1}{2}
\alpha^2 \beta^2 \left[(d_e-d_g)^2+d_{c}^2\right]^2  \right. \\ & & \hspace{-10pt} \left.
+ 2  \left[(\alpha^2 d_g + \beta^2 d_e)
(d_e-d_g) + \frac{d_{c}^2}{2}(\beta^2 - \alpha^2)\right]^2\right\}, \nonumber \\ 
\hbar C_{6}^{(2)} & =&  \frac{ \alpha^4 \beta^2 d_{c}^4}{\Delta + \sqrt{\Delta^2 +
4\Omega^2}} + \frac{\alpha^2 \beta^4  d_{c}^4}{\Delta + 3 \sqrt{\Delta^2 +
4 \Omega^2}}.\nonumber
\end{eqnarray}
For the optimal values  $d E_{\rs dc}/B = 1$ and $\Omega/\Delta = 0.9258$, the
van der Waals coefficients take the form $C_{6}^{(0)} = 0.004 \hbar \Delta
r_{\Delta}^6$ and $C_{6}^{(2)} = 0.005 \hbar \Delta r_{\Delta}^{6} $. At shorter
distances $r_{\rs B} \ll r < r_{\Delta}$, the effective interaction reduces to $
V_{\rs eff}({\bf r}) =   \left(d_{c}^2+d_{g} d_{e}\left[1-3 \cos^2
\theta\right]\right)/ r^{3} $ and remains repulsive for all angles $\theta$.
Thus it is possible to create purely repulsive interaction with large and
adjustable strength.

\begin{figure}[htb]
\begin{center}
\includegraphics[width= \columnwidth]{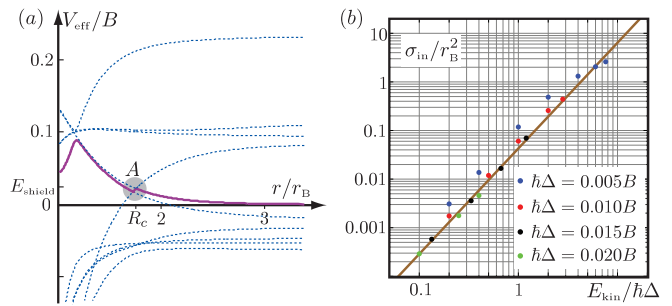}
\end{center}
\caption{(a) Born-Oppenheimer potentials for $\hbar \Delta = 0.015 B$ at $\theta
= 0$.  
The plot shows all potentials (dashed)
accessible via single-photon transitions from the state (solid) adiabatically
connected to $|+\rangle_1|+\rangle_2$. The first crossing limiting the height of
the shield appears in the region  $A$ for angles $\theta \approx 0$. 
(b) Inelastic cross section $\sigma_{\rs in}$ due
to diabatic crossings for different detunings $\Delta$ as a function of incoming
kinetic energy $E_{\rs kin}$.} \label{Fig3} \end{figure}

In order to determine the height of the potential barrier, a detailed analysis
including all internal levels is required. Such a procedure is achieved by first
deriving Born-Oppenheimer potentials accounting for the coupling of the internal
states $|j,m\rangle_{i}$ by the dipole-dipole interaction. In the second step, the
microwave field, which couples these Born-Oppenheimer potentials, 
is included within a rotating wave approximation. The new dressed levels in the rotating
frame are shown in Fig.~\ref{Fig3}(a). It follows that the height of the shield
(solid line) is limited by small avoided crossings. The first crossing (labeled
$A$) appears with the level adiabatically connected to the symmetric state
$|1,0;0,0\rangle$ for a relative orientation of the molecules along the $z$-axis
with  $\theta \approx 0$, and it limits the barrier height of the shield to $
E_{\rs shield} \approx 0.02 B$ for the optimal parameters.  The radius $R_{c}$
for the breakdown of the shield is in the range $R_{c} \sim r_{\rs B}$, which is
still large compared to the distances where additional short range interactions
have to be taken into account. This justifies our approach of restricting the
analysis to the electric dipole-dipole interaction alone.

Next, we analyze the validity of the Born-Oppenheimer approximation and study
the influence of the kinetic energy coupling different dressed potentials during
a collision. The influence of the kinetic energy is determined by the
dimensionless parameter $\gamma = d^2 m/\hbar^2 r_{\rs B}$.  For $\gamma \gg 1$,
we can apply semiclassical theory; this condition is well satisfied for typical
polar molecules like ${\rm Li Cs}$ with $\gamma \approx 6900$. For a collision
whose relative kinetic energy $E_{\rs kin}$ is  below the shield barrier, the
processes giving rise to inelastic 
loss
are (i) diabatic crossing between
different Born-Oppenheimer levels and (ii) quantum mechanical tunneling
through the barrier. We start by studying the diabatic transitions first: the
inelastic cross section within semi-classical approximation is computed by
first determining the classical trajectory $r_{\rs cl}(t)$ of a collision with a
fixed impact parameter. To determine the Landau-Zener diabatic crossings, we
solve the full Schr\"odinger equation for the internal structure using the given
relative motion $r_{\rs cl}(t)$. The loss propability is then determined by the
depletion of the adiabatic Born-Oppenheimer level at the classical turning
point.  Averaging over different impact parameters and angles of approach
provides the inelastic cross section $\sigma_{\rs in}$ due to diabatic
transitions. The main contribution comes from the Born-Oppenheimer level closely
approaching the effective potential close to $\theta \approx 0$, see
Figs.~\ref{Fig2}(a) and \ref{Fig3}(a). Note that the standard Landau-Zener
tunneling expression can not be applied here as the levels have no real
crossing.  The inelastic cross section for different $\Delta$ and $E_{\rs kin}$
is shown in Fig.~\ref{Fig3}(b). We find an algebraic behavior of the inelastic
cross section with $\sigma_{\rs in} =  \rho (E_{\rs kin}/\hbar \Delta)^{\kappa}
r_{\rs B}^2$ with $\kappa \approx 2.2$ and $\rho = 0.043$ at  $\hbar \Delta =
0.015 B$, solid line in Fig.~\ref{Fig3}(b).   The loss rate $1/\tau_{\rs in}$
during a two-particle collisions reduces to
\begin{equation}
  \frac{1}{\tau_{\rs in}} \approx 11 \frac{B \: n r_{\rs B}^{3}}{\hbar}
\left(\frac{T}{B}\right)^{\kappa +1/2}, 
\end{equation}
where we have replaced the collision energy with the temperature $T$ of the gas.
Consequently, for ${\rm Li Cs}$ at characteristic densities $n\sim 10^{13} \:
{\rm cm}^{-3}$, the lifetime is several seconds even for $T \sim 1 {\rm mK} <
\hbar \Delta$.

The second scenario for an inelastic collision is quantum mechanical
tunneling through the barrier. For $\gamma\gg 1$ such tunneling processes are
strongly suppressed and can be studied using semiclassical techniques like
WKB. The tunneling propability during a single collision is then given by the
Eucledian action for the trajectory $C$ starting at the classical
turning point $R_{0}$ and ending at the inner distance $R_{c}$, where the shield
starts to break down;
\begin{equation}
P_{\rs WKB} = \exp \left(- 2 \int_{C} d s \: \sqrt{m \left[V_{\rs
eff}({\bf r})- E_{\rs kin}\right]}/\hbar \right).
\end{equation}
Thus, the highest tunneling probability is obtained along the collision axis
with the lowest shield barrier, see discussion above.  Then the characteritic
scale for the tunneling amplitude at low incoming kinetic energies ($E_{\rs kin}
\ll \hbar \Delta$) is given by $P_{\rs WKB} = \exp(- c \sqrt{\gamma})$, where
the numerical constant for $\hbar \Delta/B=0.015$ takes the form $c\approx
0.32$.  Consequently, for ${\rm Li Cs}$  the tunneling is strongly suppressed
and can be safely ignored at low kinetic energies $E_{\rs kin} < \hbar \Delta$.

It is important to note that the qualitative behavior of the shield remains
robust during a three-body collision.  The main modification to the shield is
that the antisymmetric levels can open up an avoided crossing, as the parity
symmetry present in the two-particle problem is broken for three particles. The
relevant antisymmetric level is shown by a dotted line in Fig.~\ref{Fig2}(b).
As this crossing appears at energies $\sim \! \hbar \Delta$, it does not modify
the validity of the above discussion for incoming kinetic energies $E_{\rs kin}
< \hbar \Delta$.  Thus, the shield prevents three particles from
approaching each other on distances, where the formation of bound states can take
place, and three-body losses are therefore effectively quenched.

To summarize, we showed that properly adjusted continous wave microwave and DC
electric fields can create a repulsive shield resulting in large suppression of
inelastic collisions. We now describe possible avenues opened by this work.
For efficient evaporative cooling, it is important that elastic collisions allow
for fast thermalization. The elastic scattering cross section at low collisional
energies is determined by the $s$-wave scattering length.  In the present
situation, the scattering length can be estimated via a partial wave expansion
\cite{gribakin93}: the dominant part is determined by the isotropic part of the 
effective interaction potential $  V_{\rs eff}^{(0)}(r) =
C_{6}/r^{6}$.
%
The $s$-wave scattering length then reduces to  $a_{s} \approx \left(C_{6}
m/\hbar^2\right)^{1/4} \sim r_{\Delta} \left(d^2 m/\hbar^2
r_{\Delta}\right)^{1/4}$.  For ${\rm Li Cs}$ at the optimal detuning $\hbar
\Delta = 0.015 B$, it follows that $a_{s} \approx 66 \: {\rm nm}$ yielding a
large  elastic cross section $\sigma_{\rs el}$ with suppressed losses providing
an ideal system for evaporative cooling, e.g.\  $\sigma_{\rs el}/ \sigma_{\rs
in} \sim 10^6$ for $\sigma_{\rs in}$ at $T = 1$ mK and $\sigma_{\rs el} = 8 \pi
a_s^2$.

Another application is the creation of three-dimensional crystalline phases. The
many-body Hamiltonian for a gas of ultracold polar molecules reduces to
\begin{equation}
  H = \sum_{i} \frac{{\bf p}_{i}^2}{ 2 m} + \frac{1}{2} \sum_{i \neq j }
\frac{C_{6}}{|{\bf r}_{i}-{\bf r}_{j}|^6},
\end{equation}
where we neglect additional terms due to the anisotropy of the effective
potential and due to three-body interactions discussed in \cite{buechler07.2}.
In analogy to the appearance of crystalline phases for polar molecules confined
to two dimensions \cite{buechler07}, for strong repulsive van der Waals
interactions, the system will undergo a phase transition from a liquid phase to
a solid phase. The dimensionless parameter controlling the transition takes the
form $\lambda = C_{6} m /\hbar^2 a^4 $ with $n=1/a^3$ the particle density: for
weak interactions
$\lambda\ll 1$ the ground state is in a liquid phase, while for strong
interactions with $\lambda\gg 1$ the system is characterized by a solid phase
with broken translational symmetry. Consequently, this opens up a way towards
the creation of three-dimensional crystalline structures with ultracold
molecular gases.

We acknowledge discussions with J. Doyle and W. Campbell. The work was
supported by the European Science Foundation with EuroQUAM program, 
FWF, NSF, ARO, MURI, and DARPA.


\end{document}